\documentclass[%
12pt,
reprint,
amsfonts,amsmath,amssymb,
aps,
prb,
]{revtex4-2}

\usepackage{mathptmx}
\usepackage{color}
\usepackage[tight]{subfigure}
\usepackage{graphicx}
\usepackage{dcolumn}
\usepackage{bm}
\usepackage[colorlinks,linkcolor={blue},citecolor={blue},urlcolor={blue}]{hyperref}
\usepackage{marginnote}
\usepackage{physics}

\DeclareSymbolFont{epsilon}{OML}{cmm}{m}{it}
\DeclareMathSymbol{\epsilon}{\mathord}{epsilon}{"0F}

\newcommand{\ba}{\begin{array}}
\newcommand{\ea}{\end{array}}

\newcommand{\ra}{\rangle}
\newcommand{\la}{\langle}

\newcommand{\veps}{\varepsilon}
\newcommand{\up}{\uparrow}
\newcommand{\dn}{\downarrow}
\newcommand{\nnum}{\nonumber}

\renewcommand{\vec}[1]{\mathbf{#1}}
\newcommand{\kv}{\mathbf{k}}
\newcommand{\rv}{\mathbf{r}}

\begin{document}

\title{Restricted Boltzmann machine network versus Jastrow correlated 
wave function for the two-dimensional Hubbard model}

\author{Karthik V.}\email{karthik.v16@iisertvm.ac.in} 
\author{Amal Medhi}\email{amedhi@iisertvm.ac.in} 
\affiliation{Indian Institute of Science Education and Research Thiruvananthapuram,
Kerala 695551, India}


\begin{abstract}
We consider a restricted Boltzmann Machine (RBM) correlated BCS wave function as the ground state
of the two-dimensional Hubbard model and study its electronic and magnetic properties as a function
of hole doping. We compare the results with those obtained by using conventional Jastrow projectors.
The results show that the RBM wave function outperforms the Jastrow projected ones in 
the underdoped region in terms of the variational energy. Computation of superconducting (SC) correlations
in the model shows that the RBM wave function gives slightly weaker SC correlations as compared
to the Jastrow projected wave functions. A significant advantage of the RBM wave function is that it spontaneously gives rise to strong antiferromagnetic (AF) correlations in the underdoped region even though the wave function does not incorporate any explicit AF order. In comparison, AF correlations in the Jastrow projected wave functions are found to be very weak. These and other results obtained show that the
RBM wave function provides an improved description of the phase diagram of the model.
The work also demonstrates the power of neural-network quantum state (NQS) wave functions 
in the study of strongly correlated electron systems.
\end{abstract}

\maketitle

\section{INTRODUCTION}
The neural-network quantum state (NQS) wave functions have recently been applied
successfully to study several quantum many-body 
systems\cite{Carrasquilla_NatPhys2017, Carleo_Science2017,confusion,Broecker2017,
Carrasquilla_PhysRevX.7.031038, SDasDarma_PhysRevX.7.021021}.
The NQS wave functions are variational wave functions constructed based on an artificial neural-networks (ANNs). 
The power of such wave functions derives from the expressive capability of the ANNs which can 
represent highly nonlinear functions\cite{HORNIK1989359}. 
In contrast to the traditional Jastrow-type variational wave functions\cite{YAMAJI1998225,Gros_Edegger2007GutzwillerRVBTO,Imada_PhysRevB.96.085103} 
which are biased by the choice of a mean-field ansatz, NQS wave functions can be highly unbiased and can
provide an accurate representation of quantum many-body states as has been demonstrated in several 
works\cite{Carleo_Science2017,Neupert_PRL2018,Carleo_RevModPhys.91.045002,Carrasquilla_PhysRevX.7.031038,
Carrasquilla_AdvPhys2020,Melko2019_NatPhys2019,Carrasquilla_PRXQuantum.2.040201}.
However, most of these applications have been limited to bosonic systems. For fermionic systems, 
there is a key disadvantage that though neural-networks can represent complicated functions,
it generally fails to take into account the complicated sign structure of a 
fermionic many-body wave function\cite{Inui_PhysRevResearch.3.043126,Cai_PhysRevB.97.035116}.
Indeed, it fails to correctly represent the sign structures of the wave function of even 
non-interacting fermions\cite{Cai_PhysRevB.97.035116}.
An alternative approach in which the above problem is avoided is to use an ANN as
a correlator to an antisymmetric one-body wave function which takes care of the fermionic 
symmetry correctly. Although such a wave function is 
less unbiased by construction, it still outperforms traditional Jastrow projected 
wave functions in terms of accuracy. For example, Nomura {\em et al.}\cite{Nomura_PhysRevB.96.205152} 
have used the restricted Boltzmann machine (RBM) network as a correlator to a 
generalized pair-product wave function. The resulting RBM+PP wave function was shown to give 
significantly lower energy compared to Jastrow type wave function for the two-dimensional 
Hubbard model at half-filling. In a recent work\cite{Karthik_PhysRevB.110.125125}, the present
authors showed that a convolutional restricted Boltzmann machine (CRBM) correlated 
wave function provides a significantly improved description of the ground state of the 
half-filled Hubbard model on a square lattice.

In this work, we consider an RBM correlated BCS wave function as a ground
state of the two-dimensional Hubbard model and study it as a function of hole doping.
We examine several electronic properties like ground state energy, superconducting and magnetic 
correlations etc.\ and compare the results with those obtained by using
several standard Jastrow projectors. Specifically, the wave function we consider is of the 
form ${\cal P}\ket{BCS}_N$, where $\ket{BCS}_N$ is the BCS state
with fixed electron number and ${\cal P}$ is the projector or the correlator which introduces
electronic correlation into the wave function. We take it to be ${\cal P}_{RBM}$, a correlator 
constructed using an RBM network as well as three other Gutzwiller-Jastrow projectors.
We study the resulting wave function using the variational Monte Carlo 
method\cite{TaharaImada_VMC_JPSJ.77.114701,SorellaVMC,YokoyamaShiba_VMC1987}. We show that compared to the 
Jastrow projectors, the RBM correlated wave function gives lower energy in the underdoped
phase of the model. We obtain a ground state phase diagram of the model and show that 
the wave function give a better description of the superconducting and magnetic properties of the model.
The rest of the paper is organized as follows. We describe the model and the variational 
wave functions considered in section~\ref{sec:model_method}.  The results are described
in section~\ref{sec:results}. Finally, we give the summary and concluding remarks
in section~\ref{sec:summary}.

\section{Model and method}
\label{sec:model_method}
We consider the fermionic Hubbard model on a two-dimensional (2D) square lattice given by,
\begin{align}
\mathcal{H}=-t\sum_{\la i,j\ra\sigma}\left(c^\dag_{i\sigma}c_{j\sigma} + hc\right) 
+ U\sum_{i}n_{i\up}n_{i\dn}
\end{align}
The operator $c^\dag_{i\sigma}$ creates an electron at site `$i$' with spin $\sigma$, and $c_{i\sigma}$ is the corresponding annihilation operator.
$n_{i\sigma}=c^\dag_{i\sigma}c_{i\sigma}$ is the number operator. In the first term, the summation is over the nearest-neighbor sites. The parameter $t$ is the hopping amplitude and $U$ is the onsite Hubbard repulsive potential. Here we take $U/t=8$, a value that lies in the intermediate coupling strength regime of the model.

The RBM correlated wave function we consider is given by,
\begin{align}
|\Psi_{RBM}\rangle = {\cal P}_{RBM}\ket{BCS}_N
\end{align}
where $\ket{BCS}_N$ is the BCS ground state in the subspace of the fixed number of particles ($2N$),
of the following mean-field pairing Hamiltonian,
\begin{align}
{\cal H}_{MF} =& \sum_{\kv\sigma}\veps_{\kv}c^\dag_{\kv\sigma}c_{\kv\sigma} 
 +\sum_{\kv}\left(\Delta_{\kv}c^{\dag}_{\kv\up}c^{\dag}_{-\kv\dn}+hc\right) 
\end{align}
where $\veps_{\kv} = -2t(\cos k_x+\cos k_y)-\mu$. We take the superconducting (SC) pair 
amplitude to be $\Delta_{\kv} = \Delta_{SC}(\cos k_x-\cos k_y)$ ($d$-wave pairing).
The quantities $\Delta_{SC}$ and $\mu$ are the variational parameters in the mean-field
part of the wave function. The BCS wave function in real-space representation can be 
written as,
\begin{align}
\ket{BCS}_N = \sum_{\{i_1,\ldots i_{N},j_1,\ldots,j_N\}} 
\begin{vmatrix}
\phi(\rv_{i_1}-\rv_{j_1}) & \cdots & \phi(\rv_{i_1}-\rv_{j_N}) \\
\vdots & \cdots & \vdots \\
\phi(\rv_{i_N}-\rv_{j_1}) & \cdots & \phi(\rv_{i_1}-\rv_{j_N}) \\
\end{vmatrix}\nnum\\
 \times c^\dag_{i_1\up}\ldots c^\dag_{i_N\up}c^\dag_{j_1\dn}\ldots c^\dag_{j_N\dn}\ket{0}
\end{align}
The set $\{i_1,\ldots i_{N},j_1,\ldots,j_N\}$ represent the set of sites occupied by the electrons. The pair amplitudes $\phi(\rv_{i}-\rv_{j})$ are functions of the variational parameters $\Delta_{SC}$ 
and $\mu$\cite{Paramekanti_PhysRevB.70.054504}. 
Denoting the determinantal coefficients by $\Psi_{BCS}$, we can write,
\begin{align}
\ket{BCS}_N =& \sum_{n_1,n_2,\ldots n_{2L}}  \Psi_{BCS}(n_1,n_2,\ldots n_{2L}) \nnum\\
& \times (c^{\dag}_{1\up})^{n_1}\ldots (c^{\dag}_{L\up})^{n_L}
 (c^{\dag}_{1\dn})^{n_{L+1}}\ldots (c^{\dag}_{L\dn})^{n_{2L}}\ket{0}  \nnum\\
 =& \sum_R\Psi_{BCS}(R)\ket{R} 
\end{align}
The occupation number $n_i$-s can values $0$ or $1$. $L$ is the number of lattice sites. The
set $R\equiv \{n_1,n_2,\ldots,n_{2L}\}$ specifies the electronic configurations. 

\begin{figure}[!htb]
\centering
\includegraphics[width=0.9\columnwidth]{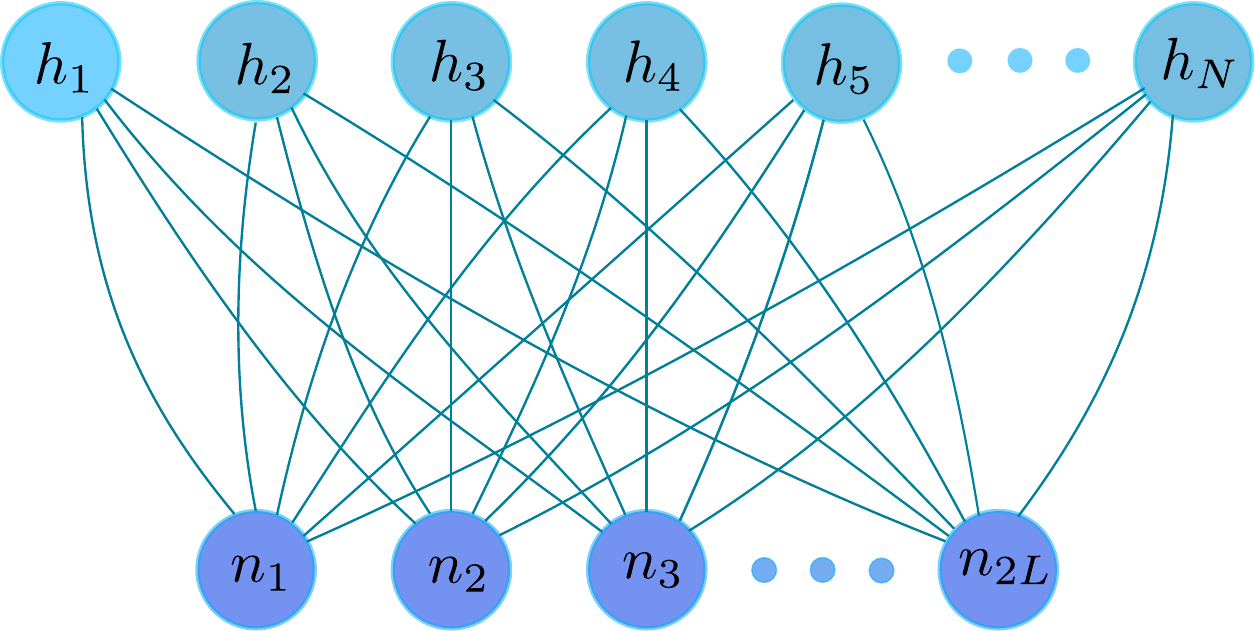}
\caption{Schematic diagram of a restricted Boltzmann machine (RBM) network. The input to the 
network are the electron occupation numbers $\{n_1,n_2,\ldots,n_{2L}\}$.
The hidden layer defines variables $h_i$ which can take values $\pm 1$.}
\label{fig:rbm_network}
\end{figure}

Now we consider a restricted Boltzmann machine (RBM) network as shown in Fig.~\ref{fig:rbm_network}.
The network consists of a visible layer with $2L$ neurons and a hidden layer with $P=2L\alpha$ 
neurons. The parameter $\alpha$ is called the hidden variable density.
Let $a_i$, $b_j$ be the bias parameters corresponding to the neurons in the visible and 
hidden layers, respectively. Also, let $w_{ij}$ be the weight parameter corresponding to the connection 
between $i$-th visible neuron and the $j$-th hidden.
For a given input configuration $R=\{n_1,n_2,\ldots,n_{2L}\}$, the RBM network defines
a function\cite{Carleo_NatComm2018},
\begin{align}
\Psi_{RBM}(R)= \frac{1}{Z}e^{\sum_ia_in_i}\prod_{j} 2\cosh\Bigl(\sum_{i}w_{ij}n_i + b_j\Bigr)
\label{eq:psi_crbm}
\end{align}
Therefore, we define the RBM variational wave function as,
\begin{align}
\ket{\Psi_{RBM}} = {\cal P}_{RBM}\ket{BCS}_N = \sum_R \Psi_{RBM}(R)\Psi_{BCS}(R)\ket{R}
\label{eq:wf_finalform}
\end{align}
where the network parameters $a_i$, $b_j$, $w_{ij}$ are treated as the variational parameters. 
The number of parameters can be greatly reduced by incorporating the symmetries of the 
Hamiltonian into the wave function. Here we use the lattice translational symmetry and
impose the condition that for two electronic configurations $R$ and $R'$ connected by 
a lattice translation operation, the corresponding wave function amplitudes are equal,
i.e.~$\Psi_{RBM}(R)=\Psi_{RBM}(R')$. This reduces the numbers of $w_{ij}$ and $b_j$ 
parameters to $4L$ and $2$, respectively. Also since we are working with fixed 
electron number representation, the sum $\sum_ia_in_i$ is a constant, and the factor 
$e^{\sum_ia_in_i}$ can be dropped from the Eq.~(\ref{eq:psi_crbm}). 
Thus total number of variational parameters becomes $(4L+2)$.

For comparison, we also obtain results for the following Jastrow 
projected wave functions. The first is the Gutzwiller (GW) projected wave function
$\ket{\Psi_{GW}} = {\cal P}_{G}\ket{BCS}_N$, where
\begin{align}
{\cal P}_G=\prod_i\left[1 - (1-g)n_{i\up}n_{i\dn}\right],\quad 0<g\leq 1 
\end{align}
The second wave function we consider is given by $\ket{\Psi_{GW+DH}} = {\cal P}_{G}{\cal P}_{DH}\ket{BCS}_N$.
It incorporates doublon-holon (DH) binding with ${\cal P}_{DH}$ given by\cite{YokoyamaJPSJ2006},
\begin{align}
{\cal P}_{DH} =& \prod_{i}\left(1 - \eta Q_i\right)  \\
Q_i =& \prod_{\delta}\left[d_i(1-h_{i+\delta}) + h_i(1+d_{i+\delta})\right]
\end{align}
where $d_i=n_{i\up}n_{i\dn}$ is doublon and $h_i=(1-n_{i\up})(1-n_{i\dn})$ is holon
operator. $\delta$ denotes the nearest-neighbor sites, and $\eta$ is a 
variational parameter. 
We also consider the long range density-density projected Jastrow wave function given by
$\ket{\Psi_{Jastrow}} = {\cal P}_{J}\ket{BCS}_N$, where
\begin{align}
{\cal P}_{J} = \exp\left(\sum_{ij}\frac{1}{2}v_{ij}(n_i-1)(n_j-1)+w_{ij}h_id_j \right)
\end{align}
Here the quantities $v_{ij}$ and $w_{ij}$ are variational parameters.

We use the variational Monte Carlo (VMC) method\cite{Ceperley_VMC_PhysRevB.16.3081,SorellaVMC,TaharaImada_VMC_JPSJ.77.114701} to compute the variational energy,
\begin{align}
E_{var}(\bm{\alpha}) = \frac{\bra{\Psi_{var}}{{\cal H}}\ket{\Psi_{var}}}{\bra{\Psi_{var}}\Psi_{var}\rangle}
\end{align} 
The wave functions are optimized by minimizing $E_{var}(\bm{\alpha})$ with respect to the 
variational parameters $\bm{\alpha}$. We use the stochastic reconfiguration (SR) technique\cite{SorellaVMC, TaharaImada_VMC_JPSJ.77.114701} for the purpose.

\section{Results}  
\label{sec:results}
\begin{figure}[!htb]
\centering
\subfigure[\label{fig1a}]{
\includegraphics[width=0.48\columnwidth]{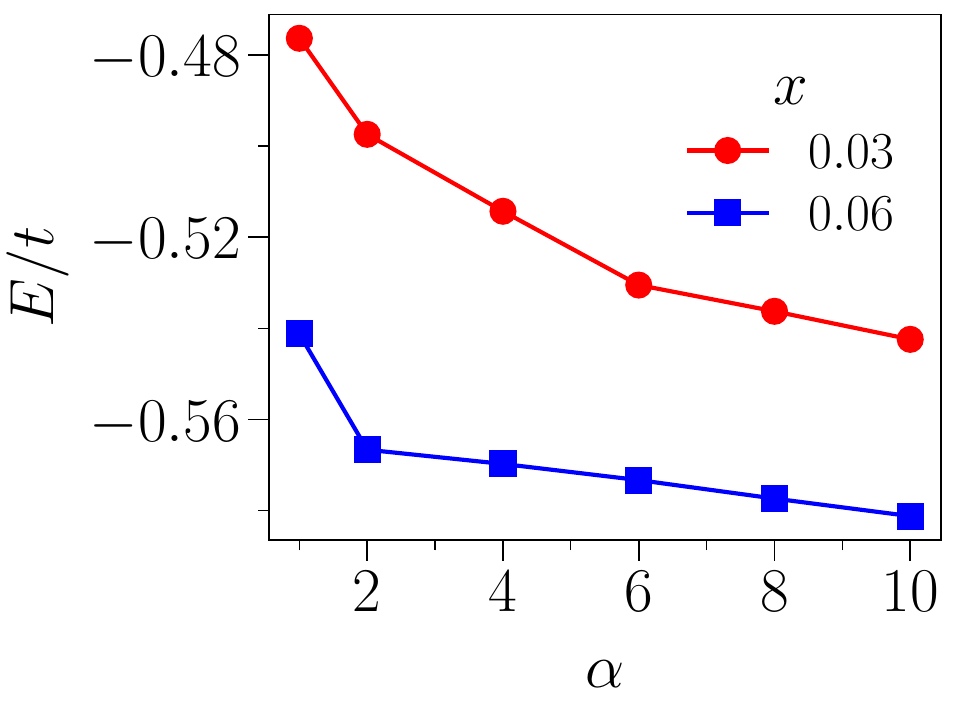}}
\subfigure[\label{fig1b}]{
\includegraphics[width=0.48\columnwidth]{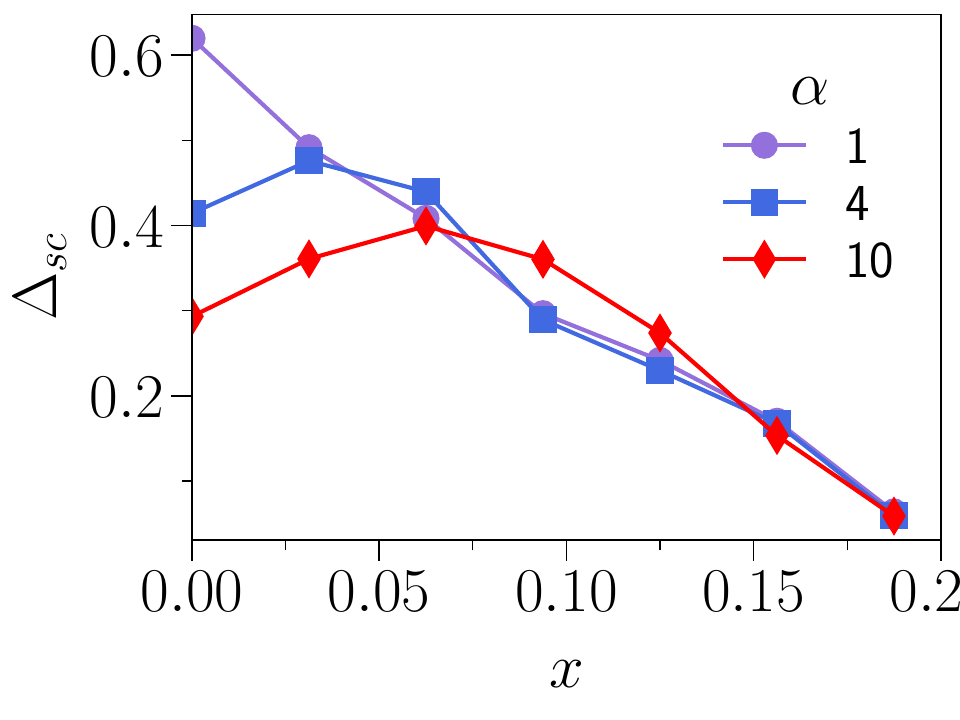}}
\caption{(a) Optimized variational energy of the RBM-SC wave function as a function of
network hidden variable density $\alpha$.  (b) Variation of optimized SC gap 
parameter $\Delta_{SC}$ with hole doping $x$ at three different values of $\alpha$. The variation
becomes dome shaped at large $\alpha$.}
\label{fig:alpha_variation}
\end{figure}

We consider a lattice of size $L=8\times 8$. The Hubbard interaction parameter is fixed at $U/t=8$. 
The RBM wave function depends upon the network structural parameter, that is the hidden variable
density $\alpha$. We need to fix the value of $\alpha$ before going ahead with the calculations.
We determine $\alpha$ by repeatedly optimizing the variational energy for a given model parameters by
taking different values of $\alpha$. The results obtained at two different hole dopings are
shown in Fig.~\ref{fig1a}. 
As the figure shows, the variational energy decreases with increasing $\alpha$ implying more
accurate wave function for larger $\alpha$.
However, increasing $\alpha$ also raises the computational complexity as the number of variational 
parameters increases rapidly with $\alpha$. Thus, one needs to make a trade-off between accuracy and 
computational efficiency. Here, we fix the parameter at $\alpha=10$ for the rest of the calculations, 
a value that gives good balance between accuracy and efficiency.
Fig.~\ref{fig1b} shows the doping dependence of the optimized SC gap parameter $\Delta_{SC}$ 
at different $\alpha$. It is interesting to note that at large $\alpha$,
where the WF becomes more accurate, the variation of optimal gap parameter $\Delta_{SC}$
as a function hole doping $x$ changes qualitatively. The variation becomes dome shaped at large $\alpha$.

\begin{figure}[!htb]
\centering
\subfigure[\label{fig:en_comp_a}]{
\includegraphics[width=0.48\columnwidth]{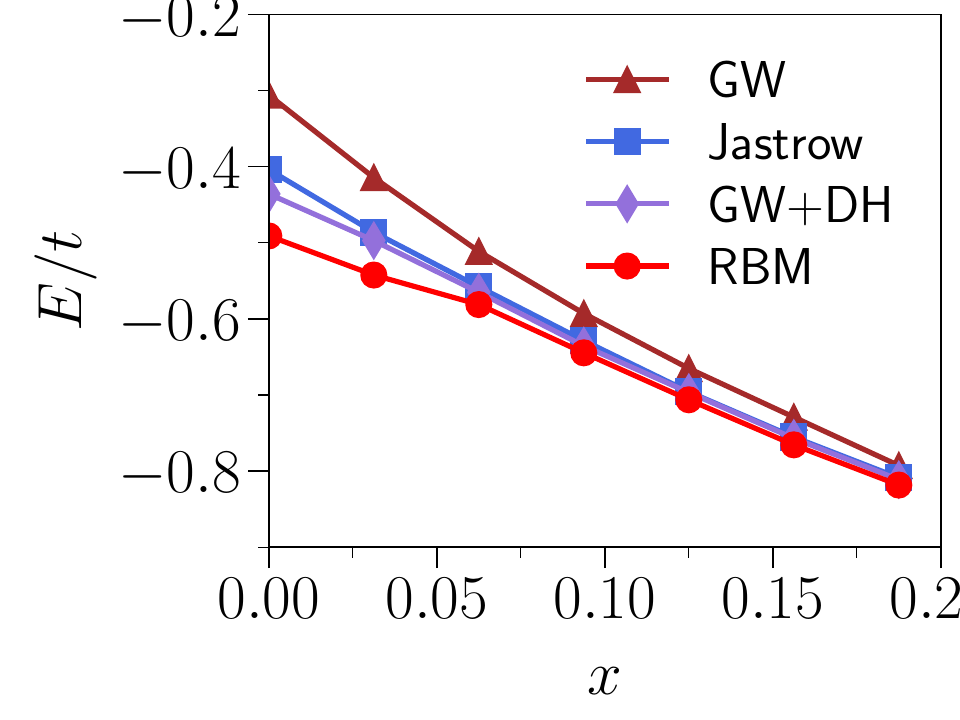}}
\subfigure[\label{fig:en_comp_b}]{
\includegraphics[width=0.48\columnwidth]{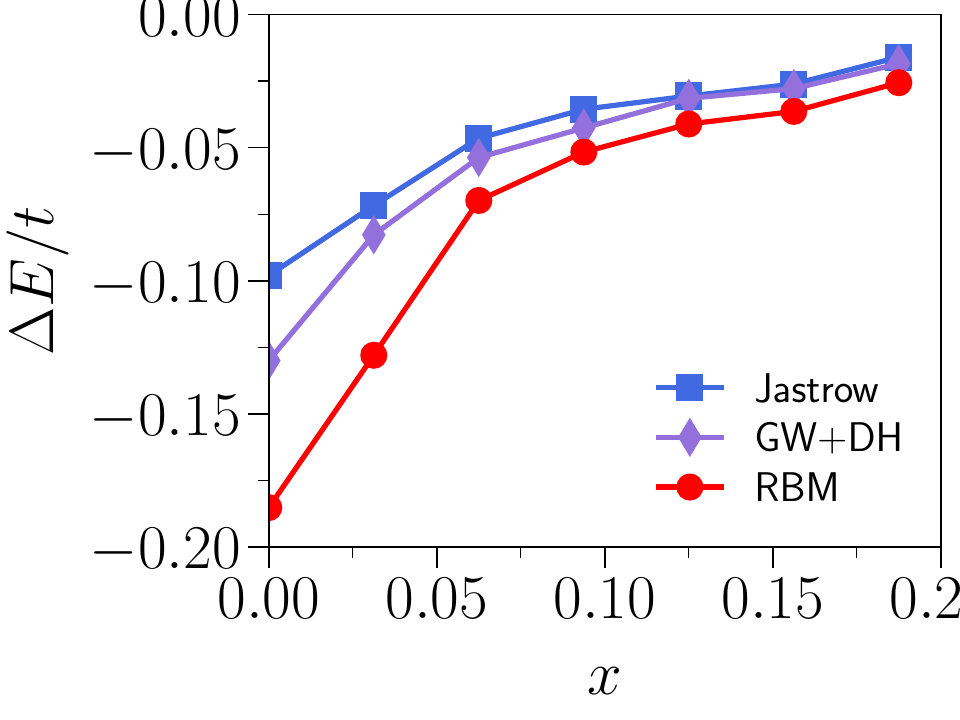}}
\caption{(a) Optimized variational energy (per site) for the four wave functions shown as a
function of hole doping $x$. (b) The energy $\Delta E/t = E_{X}/t-E_{GW}/t$ relative to
the GW energy as a function of $x$. Here $X$ stands for Jastrow or GW+DH or RBM.  $U/t=8$.}
\label{fig:en_comp}
\end{figure}

Having fixed the network parameters, we optimize the wave function as a function of hole doping $x$.
First, we examine how the energy of the RBM wave function, $\ket{\Psi_{RBM}}$ compares with those for the 
three Jastrow-type wave functions, e.g.\ $\ket{\Psi_{GW}}$, $\ket{\Psi_{GW+DH}}$, and $\ket{\Psi_{Jastrow}}$ 
defined in the previous section. The comparison of energy is shown in Fig.~\ref{fig:en_comp_a}. 
Fig.~\ref{fig:en_comp_b} shows the energy of the other three wave functions relative to 
GW energy for clarity.
As the figure shows, the GW wave function has the highest energy. As observed in 
previous studies, the simple GW projector does not take into account correlations effects adequately 
and hence is rather a poor approximation to the ground state especially for 
large $U$\cite{Kaplan_PhysRevLett.49.889,YokoyamaJPSJ2006,Capello_PhysRevLett.94.026406,
Capello_PhysRevB.73.245116}. Taking the doublon-holon (DH) binding into account improves the 
wave function substantially as energy as the curve for GW+DH shows. The energies of the GW+DH
and the long range density-density Jastrow projected WFs are significantly lower than the
GW energy. The RBM wave function gives the lowest energy and hence the most accurate among the four. 
Indeed, the RBM projector captures the correlation effect better, especially in the 
underdoped region where it also displays strong antiferromagnetic (AF) correlations as shown below.

\begin{figure}[!htb]
\centering
\subfigure[\label{fig:Sq_comp_a}]{
\includegraphics[width=0.48\columnwidth]{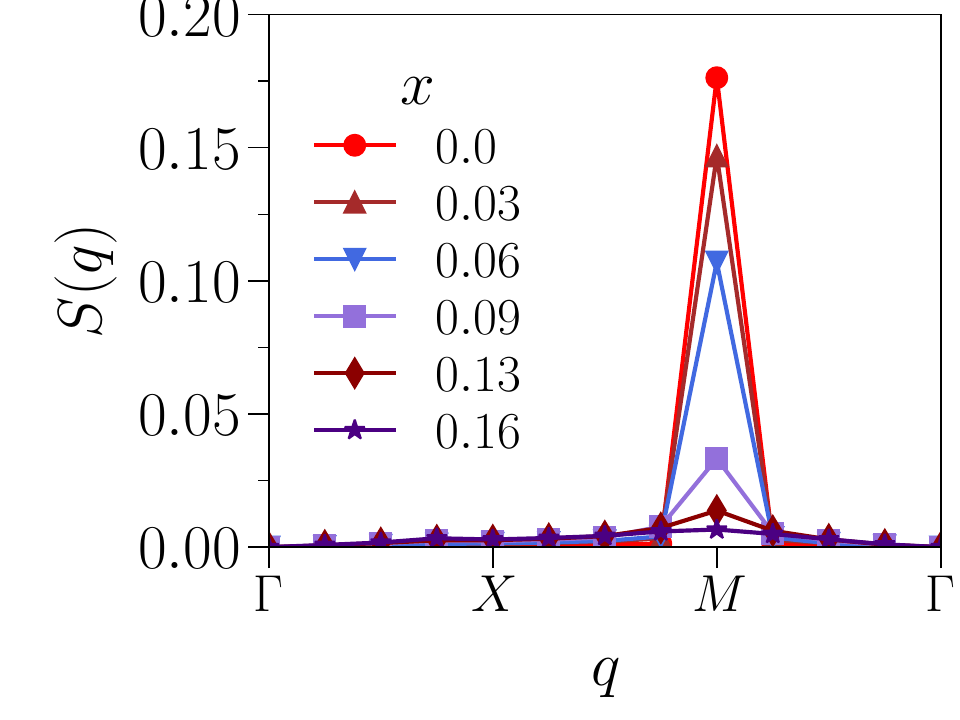}}
\subfigure[\label{fig:Sq_comp_b}]{
\includegraphics[width=0.48\columnwidth]{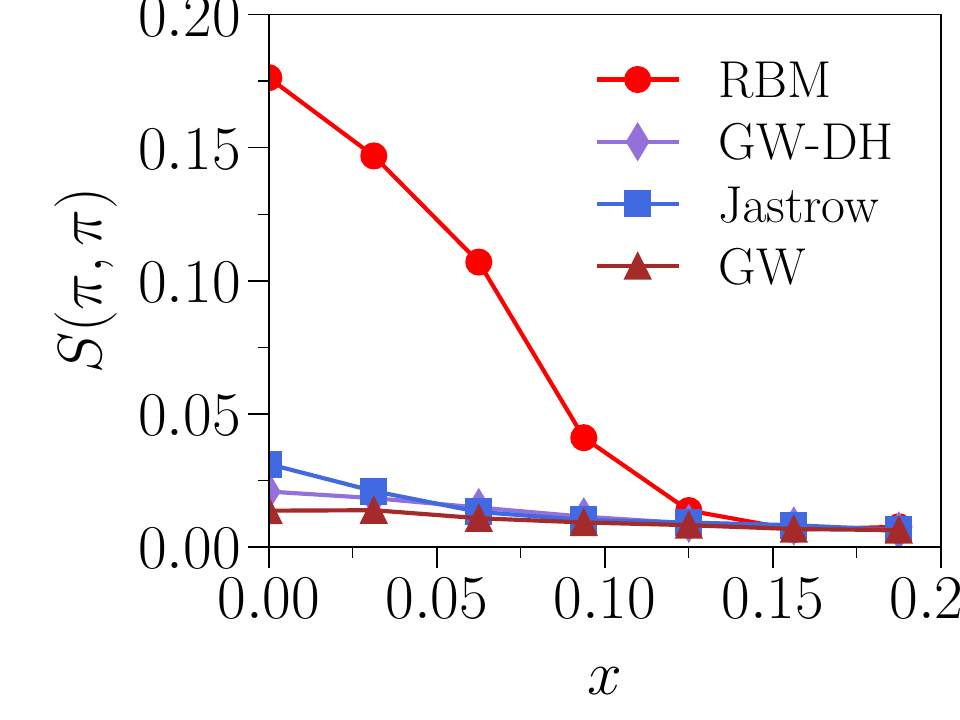}}
\caption{(a) Spin structure factor $S(\vec{q})$ versus $\vec{q}$ along the symmetry path in the
Brillouin zone of the lattice for the RBM wave function, at various values of hole doping $x$.
(b) Comparison of the staggered structure factor $S(\pi,\pi)$ at various $x$ for the four wave functions.}
\label{fig:Sq_comp}
\end{figure}

To examine the magnetic order, we calculate the spin-spin correlation 
function and spin structure factor $S(\mathbf{q})$ given by,
\begin{align}
S(\vec{q}) = \frac{1}{L^2} \sum_{i,j} e^{i\vec{q}\cdot(\rv_i-\rv_j)}\la S^z_i S^z_j \ra
\end{align}
where $S^z_i=(n_{i\up}-n_{i\dn})$ is the $z$-component of the spin operator at site $i$. 
Fig.~\ref{fig:Sq_comp_a} shows the results for $S(\vec{q})$ as a function of $\vec{q}$ along
the Brillouin zone symmetry path for the RBM wave function at various hole dopings. 
As the figure shows, $S(\vec{q})$ peaks sharply at $\vec{q}=(\pi,\pi)$ suggesting strong
AF order in the wave function, especially the underdoped region. The peak is maximum at half-filling
and drops rapidly with hole doping becoming very small for $x\gtrsim 0.1$. The strong AF order in 
the wave function is interesting given the fact that the mean-field part of the wave function,
e.g., $\ket{BCS}_N$ does not have any explicit symmetry broken AF order. This is not the 
case with the Jastrow-type projectors which give very weak AF correlations if the mean-field part 
itself does not have an explicit AF order as is the case here. This is clearly evident 
form Fig.~\ref{fig:Sq_comp_b} which shows comparison of $S(\pi,\pi)$ values for the four wave functions. 
The AF degrees of freedom in the RBM wave function can be easily understood if we look at the form
of the RBM amplitude given in Eq.~(\ref{eq:psi_crbm}). The amplitudes depends upon the matrix-vector
product $W^T \overline{n}$ which can be written as,
$\begin{bmatrix}
W_{\up} & W_{\dn} 
\end{bmatrix}
\begin{bmatrix}
\overline{n}_{\up} \\ \overline{n}_{\dn} 
\end{bmatrix}$
where $W_{\up,\dn}$ matrices are of size $2L\times L$, and $\overline{n}_{\up,\dn}$ are $L\times 1$ 
vectors representing occupation numbers of the spin-$\up$ and spin-$\dn$ states.
The elements (variational parameters) of the two sub-matrices $W_{\up}$ and $W_{\dn}$ 
are independent of each other even after the imposition of lattice translational symmetry. 
Therefore, the elements in the first and second half of the product vector $W^T \overline{n}$ are 
also distinct. Thus the weights $\Psi_{RBM}$ from the RBM network are spin-resolved by construction 
which distinguishes between different spin-configurations. These variational degrees of freedom 
make it possible to obtain AF spin configuration in the wave function upon optimization of the parameters.

\begin{figure}[!ht]
 \begin{center}
 \includegraphics[scale=0.35]{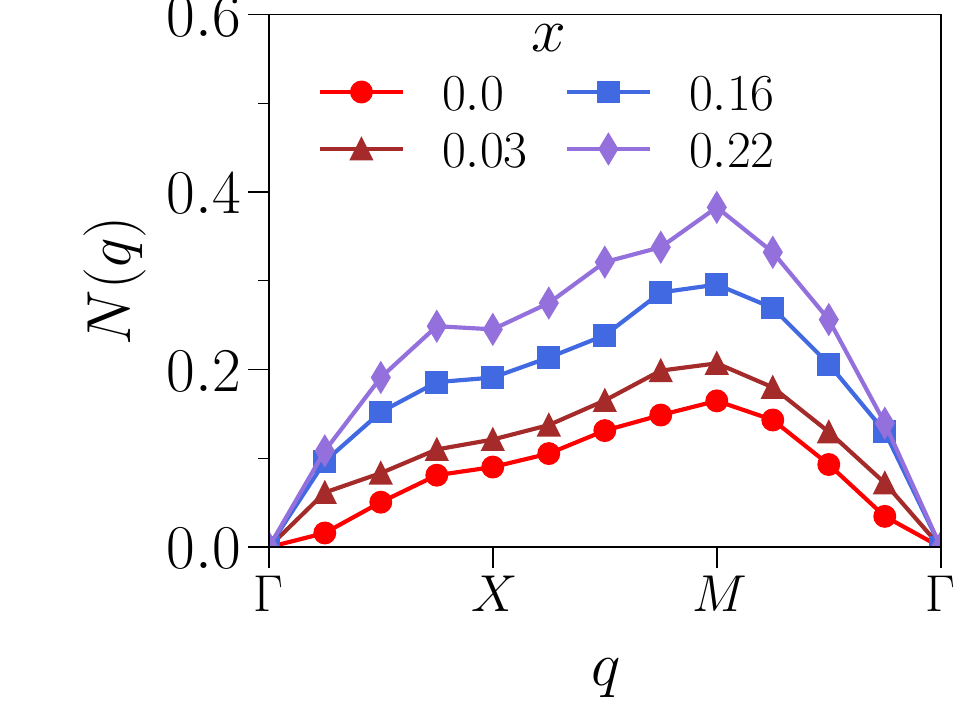}
 \end{center}
\caption{Charge structure factor $N(\vec{q})$ as a function of $|\vec{q}|$ for the RBM wave function
at various hole dopings.}
\label{fig:Nq}
\end{figure}

We calculate the charge structure factor $N(\vec{q})$ given by,
\begin{align}
N(\vec{q}) = \frac{1}{L^2} \sum_{i,j} e^{i\vec{q}\cdot(\rv_i-\rv_j)}\la n_i n_j \ra - n^{2}
\end{align}
The quantity calculated at various hole doping for the RBM-SC wave function is shown in Fig.~\ref{fig:Nq}.
As the figure shows, $N(\vec{q})$ is the smallest at half-filling. 
It increases as we increase the hole doping $x$.
At half-filling, $N(\vec{q}) \rightarrow |\vec{q}|^2$ for small $|\vec{q}|$ indicating the presence of 
charge gap. As soon as we dope the system, this behavior of $N(\vec{q})$ changes and
it goes as $N(\vec{q}) \rightarrow |\vec{q}|$, when $|\vec{q}|\rightarrow 0$. Thus the 
system becomes metallic for $x>0$.

\begin{figure}[!htb]
\centering
\subfigure[\label{fig:Phi_SC_a}]{
\includegraphics[width=0.48\columnwidth]{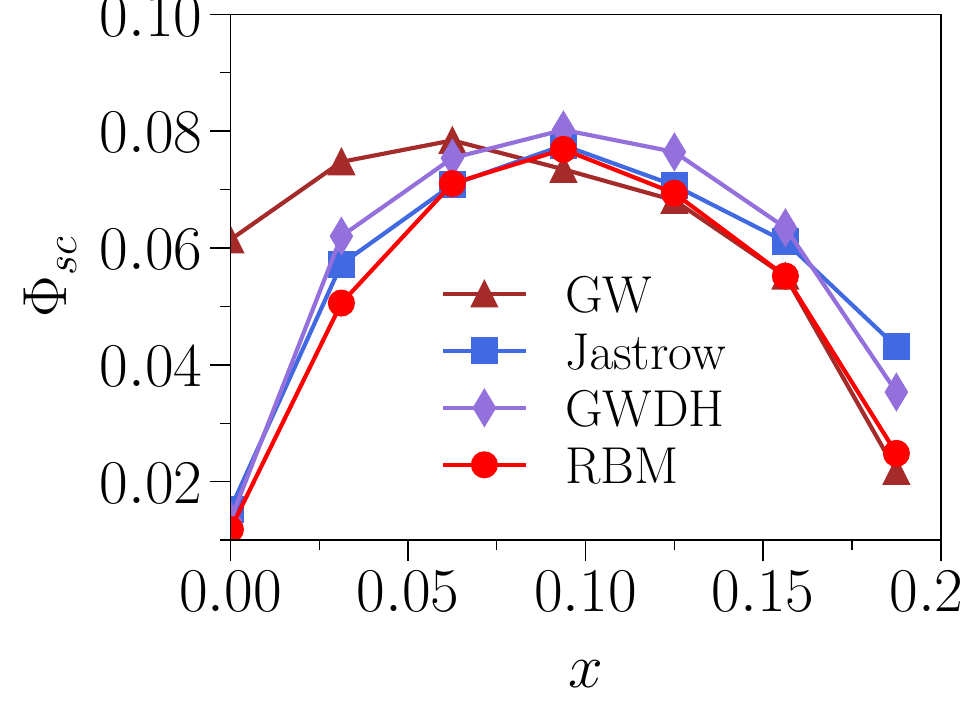}}
\subfigure[\label{fig:Phi_SC_b}]{
\includegraphics[width=0.48\columnwidth]{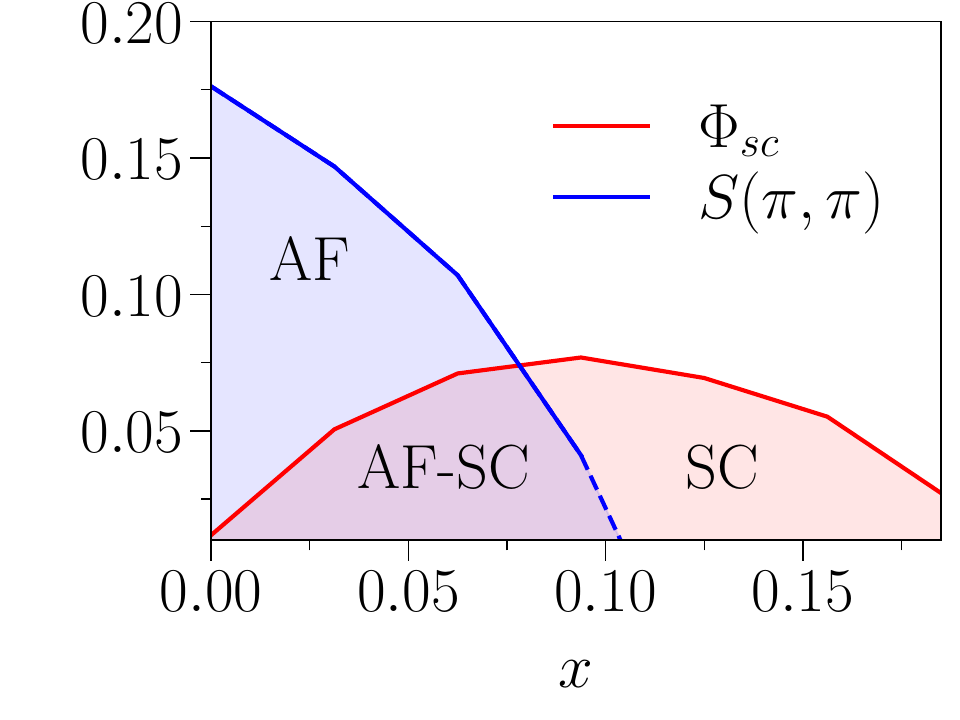}}
\caption{(a) Superconducting order parameter $\Phi_{sc}$ as a function hole doping $x$ for various
wave functions. (b) Schematic phase diagram of the model showing regions of SC and AF ordering 
as obtained from the RBM wave function.}
\label{fig:Phi_SC}
\end{figure}

Next, we calculate the superconducting pair-pair correlation function given by,
\begin{align}
F_{\alpha, \beta}\left(\mathbf{r}-\mathbf{r}^{\prime}\right)=\left\langle B_{\mathrm{r} \alpha}^{\dagger} B_{\mathrm{r}^{\prime} \beta}\right\rangle
\end{align}
where $B_{\mathrm{r} \alpha}^{\dagger}=\frac{1}{\sqrt{2}}\left(c_{\mathrm{r} \uparrow}^{\dagger} c_{\mathrm{r}+\hat{\alpha} \downarrow}^{\dagger}-c_{\mathrm{r} \downarrow}^{\dagger} c_{\mathrm{r}+\hat{\alpha} \uparrow}^{\dagger}\right)$ creates an electron singlet on the bond $(\mathbf{r}, \mathbf{r}+\hat{\alpha})$ and $\alpha$, $\beta$ 
can take values $\hat{x}$ or $\hat{y}$. We estimate the SC order parameter $\Phi_{sc}$ as 
$F_{\alpha, \beta}\left(\mathbf{r}-\mathbf{r}^{\prime}\right) \rightarrow \pm \Phi_{sc}^2$ for 
large $|\vec{r}-\vec{r}'|$, with the sign being 
$+(-)$ for $\alpha \parallel \beta(\alpha \perp \beta)$. 
Fig.~\ref{fig:Phi_SC} shows the order parameter as a function hole doping for various wave functions, 
which shows the familiar dome shaped variation of $\Phi_{sc}$ with $x$.
The GW wave function clearly overestimates the strength of SC correlations in the underdoped region,
even giving a non-vanishing $\Phi_{sc}$ at half-filling. This is obviously due to the failure of the GW projector to take into account electron correlation effects adequately. 
For the other three wave functions, $\Phi_{sc}$
is roughly of similar magnitudes, except for in the underdoped region where $\Phi_{sc}$ 
is marginally on the lower side for the RBM wave function. As described above, the RBM wave function
also displays strong AF correlations in the underdoped region whereas it is very weak for the other
wave functions. Given the fact that $\Phi_{sc}$ is roughly the same for these wave functions, it follows
that the SC correlations are unaffected by the presence of AF order.
Having obtained the value of SC and AF order parameters in the RBM wave function, we obtain a 
superconducting phase diagram as a function of hole doping which is shown in Fig.~\ref{fig:Phi_SC_b}.
As depicted in the figure, the ground state is AF insulating at half-filling. Away from half-filling,
it becomes superconducting with $\Phi_{sc}$ becoming maximum at around $x\sim 0.1$. 
Antiferromagnetism coexists with superconductivity in the underdoped region. $\Phi_{sc}$ decreases
gradually in the overdoped region and vanishes at around $x\sim 0.2$. The phase diagram 
qualitatively agrees with the results found by other studies that include the AF order in 
the mean-field state\cite{Paramekanti_PhysRevB.70.054504,YokoyamaJPSJ2006,Tocchio_PhysRevB.94.195126}. 

\begin{figure}[!ht]
\begin{center}
\includegraphics[width=0.9\columnwidth]{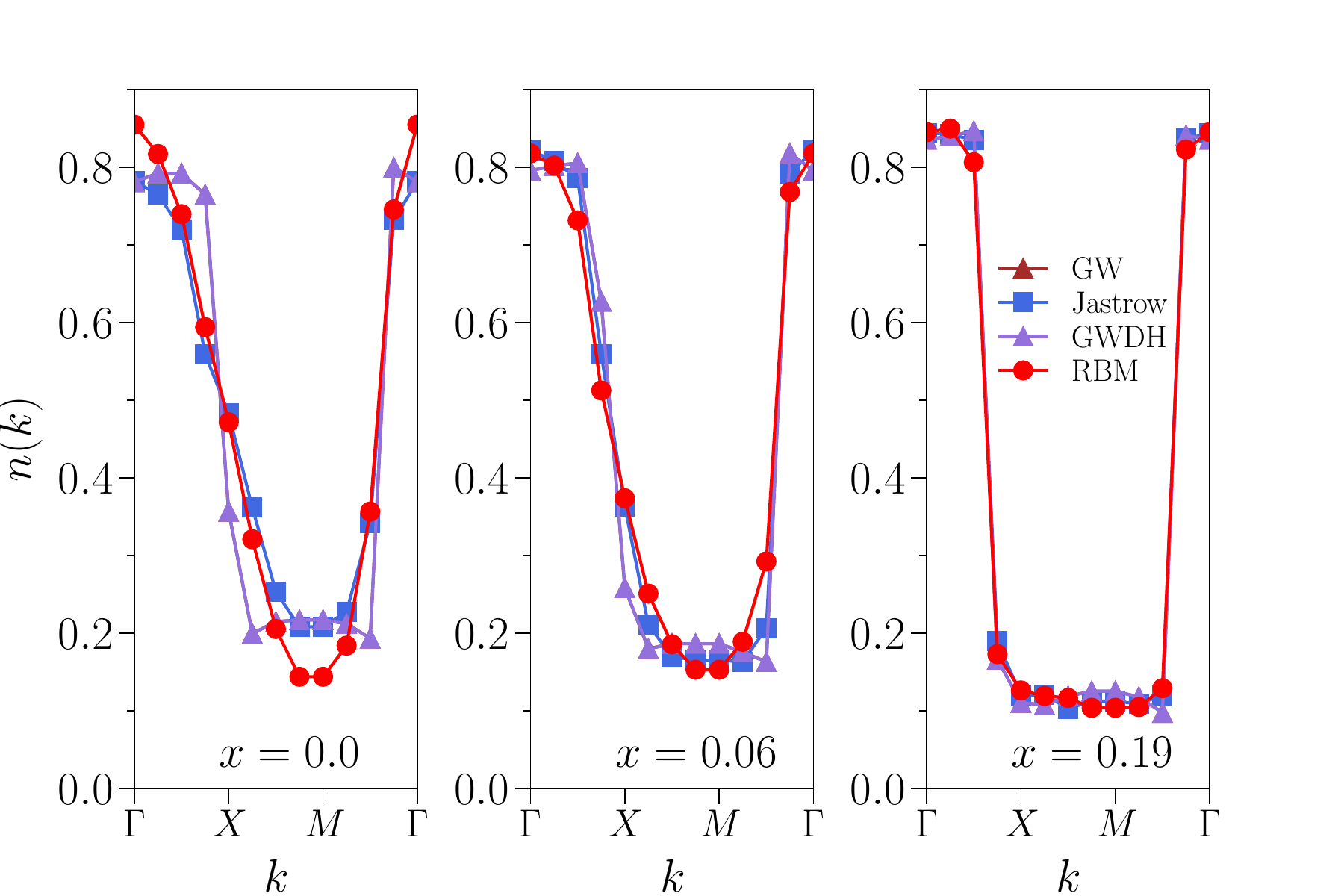}%
\end{center}
\caption{\label{fig:nk1} Momentum distribution $n(\kv)$ for the four wave functions shown in figure
as a function of $\kv$ along the symmetry path in the Brillouin zone, at three different 
values of hole doping $x$.}
\end{figure}

Finally, we look at the momentum distribution function $n(\vec{k})=\la c^{\dagger}_{\vec{k} \sigma} c_{\vec{k} \sigma}\ra$ in the wave functions. 
The values of $n(\vec{k})$ for the four wave functions are shown in Fig.~\ref{fig:nk1} for three different 
hole dopings. Since our lattice size and hence the number of data points are rather small, it is 
difficult to make an accurate comparison. Nevertheless, the results suggest that at $x=0$,
the $n(\vec{k})$ curve for the RBM wave function is a smooth one without having any discontinuity 
as it should be for a Mott insulating state. The curves for the Jastrow wave functions are somewhat
qualitatively different. Here the shift in $n(\kv)$ weights from the Brillouin zone center, $\Gamma$ 
to zone center $M$ is somewhat more stronger. At $x=0.06$ in the underdoped region, 
$n(\kv)$ for the Jastrow wave functions shows a clear jump at the Fermi wave vector in the 
nodal $\Gamma-M$ direction. In contrast, the curve for the RBM wave function remains smooth signifying
non-Fermi liquid nature of the state in the underdoped region. At the overdoped value $x=0.19$,
the curves for all the wave functions merge with each other showing once again that the wave functions
become equivalent in this region.

\section{Summary}
\label{sec:summary}
In summary, we have considered an NQS wave function obtained by using the RBM network as a correlator 
to the BCS wave function and used it to study the ground state of the Hubbard model as a function 
of hole doping. The wave functions considered are of the form ${\cal P}\ket{BCS}_N$, where ${\cal P}$ 
is a projection operator and $\ket{BCS}_N$ is the BCS wave function of an underlying mean-field 
Hamiltonian. Different choices of the projector ${\cal P}$ we considered are ${\cal P}_{RBM}$, ${\cal P}_{GW}$, 
${\cal P}_{GW+DH}$, and ${\cal P}_{Jastrow}$. The first one is constructed using an RBM network whereas the last 
three are conventional Jastrow projectors.
We examined several properties of the model using these wave functions and made a detailed comparison
among them. We found the RBM correlated wave function to be most accurate in terms of ground 
state energy as compared to the Jastrow projected ones, especially in the underdoped region of the model. 
At half-filling, all the wave functions except for the GW projected one describes a Mott insulating state. 
Superconductivity appears away from half-filling. The most interesting feature is that while all the 
Jastrow projected wave function describes a weak antiferromagnetic (AF) order, the AF corrections
in the RBM wave function is found to be very strong, even though the mean-field part of the wave function
does not contain any explicit magnetic order. This occurs as the RBM network naturally generates
spin-resolved weights by construction. Overall, the RBM projector offers an improved description 
of the ground state.
On the flip side, the RBM network though is a computationally efficient neural-network framework, it 
still involves significantly higher computational difficulty, especially in optimization, as the number 
of variational parameters grows rapidly with the system size. Therefore, it is desirable to explore other 
network architectures for better computational efficiency and significantly higher accuracy. We 
considered the convolutional restricted Boltzmann machine (CRBM) network which was found to be 
much more accurate and computationally efficient for the model at 
half-filling\cite{Karthik_PhysRevB.110.125125}, but encountered issues with numerical stability 
during optimization away from half-filling.

\section*{Acknowledgement} 
The authors thank Anusandhan National Research Foundation (ANRF), Govt of
India for financial support under the Core grant (No: CRG/2021/005792). 
Also acknowledge CHPC, IISER Thiruvananthapuram for computational facilities. 

\bibliography{references}





\end{document}